\documentclass[aip,jcp,reprint,superscriptaddress,citeautoscript,longbibliography,floatfix]{revtex4-1}

\usepackage[latin9]{inputenc}
\usepackage{amsmath}
\usepackage{amssymb}
\usepackage{graphicx}

\usepackage{enumerate}

\renewcommand{\epsilon}{\varepsilon}
\renewcommand{\epsilon}{\varepsilon}
\newcommand{\figurewidth}{1.0\columnwidth}
\newcommand{\narrowfigurewidth}{0.6\columnwidth}

\begin{document}

\title{Asymmetric Electrolytes near Structured Dielectric Interfaces}

\author{Huanxin Wu}
\affiliation{Department of Physics and Astronomy,
    Northwestern University, Evanston, Illinois 60208, U.S.A.}
\author{Honghao Li}
\affiliation{Department of Materials Science and Engineering,
    Northwestern University, Evanston, Illinois 60208, U.S.A.}
\author{Francisco J. Solis}
\affiliation{School of Mathematical and Natural Sciences, 
    Arizona State University, Glendale, Arizona 85069, U.S.A.}
\author{Monica Olvera de la Cruz}
\affiliation{Department of Materials Science and Engineering,
    Northwestern University, Evanston, Illinois 60208, U.S.A.}
\affiliation{Department of Chemistry,
    Northwestern University, Evanston, Illinois 60208, U.S.A.}
\affiliation{Department of Physics and Astronomy,
    Northwestern University, Evanston, Illinois 60208, U.S.A.}
\affiliation{Department of Chemical and Biological Engineering,
    Northwestern University, Evanston, Illinois 60208, U.S.A.}
\author{Erik Luijten}
\affiliation{Department of Materials Science and Engineering,
    Northwestern University, Evanston, Illinois 60208, U.S.A.}
\affiliation{Department of Engineering Sciences and Applied Mathematics,
    Northwestern University, Evanston, Illinois 60208, U.S.A.}
\affiliation{Department of Physics and Astronomy,
    Northwestern University, Evanston, Illinois 60208, U.S.A.}
\email{luijten@northwestern.edu}

\begin{abstract}
  The ion distribution of electrolytes near interfaces with dielectric
  contrast has important consequences for electrochemical processes and many
  other applications.  To date, most studies of such systems have focused on
  geometrically simple interfaces, for which dielectric effects are
  analytically solvable or computationally tractable.  However, all real
  surfaces display nontrivial structure at the nanoscale and have, in
  particular, nonuniform local curvature.  Using a recently developed, highly
  efficient computational method, we investigate the effect of surface
  geometry on ion distribution and interface polarization.  We consider an
  asymmetric 2:1 electrolyte bounded by a sinusoidally deformed solid
  surface. We demonstrate that even when the surface is neutral, the
  electrolyte acquires a nonuniform ion density profile near the surface. This
  profile is asymmetric and leads to an effective charging of the surface. We
  furthermore show that the induced charge is modulated by the local
  curvature. The effective charge is opposite in sign to the multivalent ions
  and is larger in concave regions of the surface.
\end{abstract}

\maketitle

\section{Introduction}
\label{sec:intro}

The behavior of electrolytes near interfaces has important consequences for
the properties of surfaces and for processes that take place in their
vicinity, such as redox reactions in electrochemical capacitors\cite{simon08},
ion transfer at biomembranes\cite{shirai95}, controlling the surface tension
of aqueous solutions\cite{wagner24,onsager34}, and establishing colloidal
stability via electric double layers\cite{hansen00}.  Despite being at the
very foundation of modern electrochemistry, complete understanding of
electrolyte structure at interfaces is still elusive.  Direct experimental
probes of the electrolyte structure near an interface have long been
challenging\cite{grahame61,favaro16}.  Theoretical approaches have used the
classical Poisson--Boltzmann (PB) model, which offers a good description for
dilute symmetric electrolytes, but often breaks down at high concentrations,
in asymmetric electrolytes, or near strongly charged
surfaces.\cite{rouzina96,belloni98,levin02} This breakdown is due to features
ignored in the mean-field model, such as finite ion
size~\cite{stern24,andresen04}, hydration shells~\cite{randles77}, dielectric
effects~\cite{wagner24}, and the molecular-scale structure of the liquid
solution~\cite{luo06}.  Many refinements in the theory have been made,
including the modified PB equation\cite{borukhov97}, the
Born--Green--Yvon\cite{henderson83} and the hypernetted chain
approximation~\cite{lozada82}, charge renormalization~\cite{ding16}, and the
inclusion of ionic polarization~\cite{levin09}.  However, we are still far
from a complete description.

Surface structure can have a strong influence on interfacial properties.
In fact, physical roughness should be carefully considered in many
applications~\cite{pajkossy94,daikhin96,vrijenhoek01}.  For example, the
Derjaguin--Landau--Verwey--Overbeek (DLVO) interaction, determined by the
repulsive double layer and the attractive van der Waals interaction, is 
significantly different for rough surfaces than for perfectly smooth
ones~\cite{bhattacharjee98,bowen00,bowen02,bradford13}.  Moreover, due to the
permittivity mismatch at the interface, ions induce polarization charges on
the surface, which are nontrivial for surfaces with nonzero curvature.

Numerical solutions to the polarization problem offer a possible path to the
investigation of these structured interfaces.  However, even with the rapid
growth of computational power, previous simulation studies have primarily
focused on geometrically simple surfaces, where the method of image charges or
other techniques can be
exploited~\cite{messina02a,messina02,dossantos11,lue11,jing15,antila18a}.  For
structured interfaces one can resort to finite-difference or finite-element
methods. Such algorithms involve discretization of the full three-dimensional
space, while (for systems with piecewise uniform permittivity) the induced
charges only reside at the interface.  Thus, these methods are inefficient for
dynamic simulation purposes, which require updating the polarization charges
at each time step.  Recently, boundary element method (BEM)-based approaches
have gained popularity~\cite{barros14b}. Unlike volume discretization methods,
the BEM only discretizes the interface and solves the Poisson equation
directly to obtain the polarization charges, which then can be readily
utilized in molecular dynamics (MD) simulations.  In this paper, we apply the
iterative dielectric solver (IDS), a recently developed fast BEM-based
dielectric algorithm optimized for MD simulations,\cite{barros14a,wu18a} that
has been demonstrated to be competitive to image-based methods~\cite{gan15},
to study the structured interfaces.  The IDS has been applied to study
complicated dielectric geometries such as patchy colloids\cite{wu16,han16a}
and self-assembly in binary suspensions.\cite{barros14b}

The surface structures that are of interest have nanoscale dimensions, making
first-principle or all-atom simulations infeasible.  We therefore employ a
coarse-grained model with implicit solvent, which captures excluded-volume
effects effects, ionic interactions and concentration fluctuations, and the
polarization effects.  To focus on the dielectric effects, we study neutral
dielectric interfaces, where the electrostatic interaction between the
interface and the ions is purely due to surface polarization charges.  To
complement the simulations we analytically study the same system to determine
the contribution to the electric potential by charges near the interface due
to their interaction with the surface.  This calculation identifies the origin
of the charge accumulation at the surface and its dependence on curvature.

\section{Model and method}

We consider a neutral solid--liquid interface $\mathbf{S}$, separating a solid
with uniform relative permittivity~$\epsilon_\textrm{s}$ from a liquid with
uniform relative permittivity~$\epsilon_\textrm{m}$.  The surface is assumed
smooth so that an outward (i.e., pointing to the liquid phase) normal unit
vector $\hat{\mathbf{n}}(\mathbf{s})$ is defined at each surface
point~$\mathbf{s} \in \mathbf{S}$.  The surface polarization charge
density~$\sigma(\mathbf{s})$ satisfies the boundary equation
\begin{equation}
  \bar{\epsilon}\sigma(\mathbf{s}) + 
  \epsilon_0\Delta\epsilon\hat{\mathbf{n}}(\mathbf{s})
  \cdot\mathbf{E}(\mathbf{s}) = 0\ ,
\label{eq:bem}
\end{equation}
where $\epsilon_0$ is the vacuum permittivity,
$\bar{\epsilon} = (\epsilon_\mathrm{s} + \epsilon_\mathrm{m})/2$ is the mean
relative permittivity, and
$\Delta\epsilon = \epsilon_\mathrm{m} - \epsilon_\mathrm{s}$ is the
permittivity contrast.  The electric field~$\mathbf{E}(\mathbf{s})$ includes
contributions from all free and bound charges, i.e., the ions in the liquid
medium as well as the surface polarization charge. Explicitly,
\begin{eqnarray}
  \mathbf{E}(\mathbf{s}) &=&
  \lim_{\delta\to 0}\iint\displaylimits_{\mathbf{S},|\mathbf{s}-\mathbf{s}'|>\delta}
  \frac{\sigma(\mathbf{s}')(\mathbf{s}-\mathbf{s}')}%
  {4\pi\epsilon_0|\mathbf{s}-\mathbf{s}'|^3} d\mathbf{s}' \nonumber\\
  && + \iiint\displaylimits_{\mathbf{V}\setminus\mathbf{S}}
  \frac{\rho_{\mathrm{f}}(\mathbf{r}')(\mathbf{s}-\mathbf{r}')}
  {4\pi\epsilon_0\epsilon_\mathrm{m}|\mathbf{s}-\mathbf{r}'|^3} 
  d\mathbf{r}' \;,
\label{eq:efield}
\end{eqnarray}
where an infinitesimal disk $|\mathbf{s}-\mathbf{s}'| \leq \delta$ is excluded
to avoid divergence of the layer potential.  $\rho_{\mathrm{f}}(\mathbf{r}')$
is the bulk free charge density in the liquid, representing the ion
distribution.  Combined, Eqs.\ (\ref{eq:bem}) and~(\ref{eq:efield}) result in
an integral equation for the surface charge that must to be solved
self-consistently.  Analytical solutions only exist for simple geometries, but
numerical solution is possible for general geometries via surface
discretization.  We assume that across each discretized surface element the
surface charge density is distributed uniformly and use a one-point quadrature
to evaluate Eq.~\eqref{eq:efield}.  Equation~\eqref{eq:bem} can then be
transformed into a matrix equation, which in the IDS is solved efficiently via
a combination of a fast Ewald summation method and the generalized minimal
residual (GMRES) method; see Ref.~\citenum{barros14a} for a detailed
description of the IDS\@.  Addressing the interactions of the ions with the
dielectric interface in this manner allows us to carry out simulations of a
primitive model in this environment. We model the hydrated ions as equisized
spheres of diameter $\sigma = 7.14$~\AA\ with point charges of valence $Z_i$
embedded at their centers.

\begin{figure}
  \centering
  \includegraphics[width=\narrowfigurewidth]{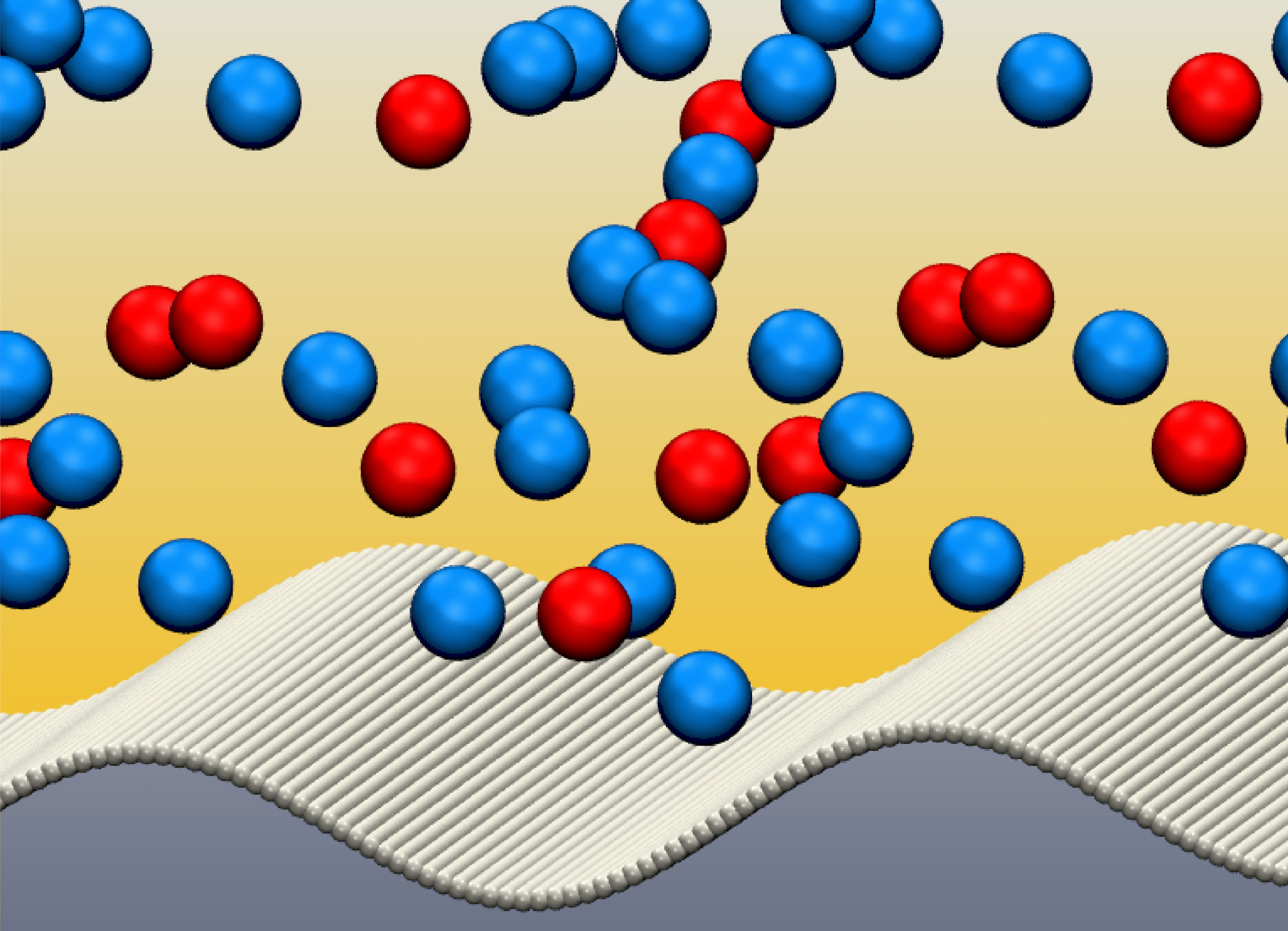}
  \caption{Primitive model of an asymmetric aqueous electrolyte near a neutral
    sinusoidal dielectric interface. A 2:1 salt solution is represented by
    positive divalent ions (red) and negative monovalent ions (blue) immersed
    in a continuous medium with relative permittivity
    $\epsilon_{\mathrm{m}} = 80$.  The medium below the (impenetrable)
    interface has relative permittivity $\epsilon_{\mathrm{s}} = 2$. Apart
    from the polarization charges, the ions also interact with the surface via
    excluded-volume interactions represented by a shifted-truncated
    Lennard-Jones potential.}
\label{fig:system}
\end{figure}
 
In nature, biomolecular structures, such as membranes and proteins, often
display complicated surface morphology.  Likewise, synthetic materials are
often not perfectly planar. As a first model, we consider a solid--liquid
interface with sinusoidal surface topography (Fig.~\ref{fig:system}).  The
system is considered as piecewise uniform, with a liquid electrolyte above a
low-permittivity solid medium. We use $\epsilon_{\mathrm{s}} = 2$ for the
solid, representing materials such as lipid
bilayers\cite{mashaghi12,dilger79,tian10}.  The local height of the
solid--liquid interface is described by $z = A\cos(2\pi x/\lambda)$, where $A$
is the amplitude of the height oscillation and $\lambda$ its wavelength.  We
start our discussion with a configuration with $A = \sigma$ and
$\lambda = 10\sigma$; below we will vary the amplitude and the surface
structure. Since the length scale of the ``roughness'' of our surface is much
larger than the size of a water molecule, we treat the background solvent as
an implicit dielectric continuum of relative permittivity
$\epsilon_{\mathrm{m}} = 80$.  The interface is discretized into a curved
rectangular mesh. To capture the excluded-volume effects and the atomistic
nature of the surface, each mesh point interacts with the ions via a
shifted-truncated Lennard-Jones (LJ) interaction. The distance between
adjacent mesh points is $0.2\sigma$, resulting in a mesh of $53 \times 49$
elements (at amplitude $A=\sigma$; more at larger amplitude).  Such a fine
mesh also guarantees an error less than $10^{-3}$ in the force calculation of
the IDS for worst-case configurations, i.e., when ions approach the surface
most closely.

For symmetric electrolytes, polarization charges induced by negative and
positive ions must cancel on average.  However, for asymmetric electrolytes
more interesting effects may occur.  We focus on 2:1 electrolytes at a
representative concentration of 50~mM. The simulation cell is periodic in the
$x$ and $y$ directions, with lateral dimensions $10\times10 \sigma^2$. The
cell is nonperiodic in the $z$ direction with height $100\sigma$, extending
from $z=-50\sigma$ to $z=50\sigma$.
The dielectric interface is centered at $z = 0$, so that the electrolyte only
resides in the upper half of the box. This slab height is sufficient to
eliminate boundary effects.  The temperature~$T$ is controlled via a Langevin
thermostat with damping time $20t_0$, where
$t_0 = (m\sigma^2/k_\mathrm{B}T)^{1/2}$ is the unit time, with
$k_{\mathrm{B}}$ Boltzmann's constant and $m$ the ion mass. The system is kept
at room temperature, so that the Bjerrum length $l_{\mathrm{B}} = \sigma$. All
simulations are performed with a time step $0.1t_0$.  The relative accuracy of
the Ewald summation is $10^{-4}$ and periodicity effects in the $z$ direction
are suppressed via a vacuum layer of $200\sigma$ and a dipole correction.  For
each parameter choice we performed independent $1600$ runs of $90\,000$ time
steps each, corresponding to $1600 \times 600$ independent samples. In the
presence of dielectric effects, we chose runs that were $4$ times shorter,
owing to the computational cost involved.

\section{Mean-field model}

To better understand the features observed in the simulations presented below,
we analyze the properties of solutions to Eq. \eqref{eq:bem} for a single ion
of valence $Z$ near the interface.  These results indicate the dependence of
induced charges on curvature.  The solution is obtained as a perturbative
expansion in the surface amplitude~$A$.  Once the induced polarization charge
is determined, the excess energy of the ion due to the polarization effects
follows as $U=Ze\phi_{\mathrm{P}}/2$, where $\phi_{\mathrm{P}}$ is the
electric potential due only to the polarization charges.  The Boltzmann weight
$\exp(-U/k_\mathrm{B}T)$ is then used to determine the relative depletion of
ions at the interface.

The perturbative approach expands the polarization potential as
$\phi_\mathrm{P}=\phi_\mathrm{P}^{(0)}+\phi_\mathrm{P}^{(1)}+\ldots,$ and
similar expansions are applied to the charge density and the geometric
quantities. The order of a term in the expansion is the power of the
modulation amplitude $A$ that appears in the expression.  The zeroth order of
this calculation corresponds to the case of a single ion near a flat
interface. In that case, we have
$\bar{\epsilon}\sigma^{(0)} +
\epsilon_0\Delta\epsilon\hat{\mathbf{n}}^{(0)}\cdot \mathbf{E}^{(0)} = 0$.
In this flat geometry, the normal component of the electric field
$\mathbf{E}^{(0)}\cdot \hat{\mathbf{n}}^{(0)}$ is due only to the single bulk
ion. Each surface polarization charge in the first term on the right-hand side
of Eq.~\eqref{eq:efield} produces (at any other point on the surface) a field
parallel to the surface and does not contribute to the normal component. Note
that the integral excludes a small region around the evaluation point, so that
a surface charge located at the point of evaluation does not contribute to the
electric field.  Thus, we have
$\mathbf{n}^{(0)}\cdot\mathbf{E}^{(0)}=Ze(4\pi
\epsilon_0)^{-1}(\mathbf{x'-x})\cdot\mathbf{n^{(0)}}/|\mathbf{x}'-\mathbf{x}|^3$,
where $\mathbf{x}'$ is a point at the surface and we take the ion position as
$\mathbf{x}=(0,0,a)$.  Integration of the Coulomb potential due to the
resulting surface charge density gives the standard image-charge potential at
the position of the ion:
$\phi_\mathrm{P}^{(0)}=(4\pi\epsilon_0)^{-1}[\Delta
\epsilon/(2\bar{\epsilon})] Ze/(2a)$.
The resulting energy of the ion is
$U^{0}=(4\pi\epsilon_0)^{-1}[\Delta \epsilon/(2\bar{\epsilon})] Z^2e^2/(4a)$.
This expression is positive when the solid phase has a lower permittivity.

The first-order term in the expansion of the potential is associated with
deformation of the surface. To simplify its calculation, we consider the limit
where the ion is brought toward the interface. In addition, we first consider
the case where its position coincides with a peak of the deformed
surface. Results for other positions follow from this calculation.  According
to the image-charge result, the energy in this limit is singular but the
exclusion of a small region around the ion renders the result finite.  We take
the excluded region as spherical, with radius $a$. That is, we use the
original distance of the ion to the surface as the cutoff for the divergent
terms. This choice is not essential but simplifies the presentation of the
results. The evaluation of the potential retains an explicit dependence on the
wavelength, which is the key feature of interest in our analysis.  A more
complex calculation, maintaining the ion at a finite distance from the
interface, gives similar results. In this limit, the first-order terms in
Eq.~\eqref{eq:bem} read
$\bar{\epsilon}\sigma^{(1)}+\epsilon_0\Delta\epsilon\hat{\mathbf{n}}^{(1)}\cdot\mathbf{E}^{(0)}=0$. 
Other terms in the expansion of the equation cancel owing to the geometry used.
The first-order term in the expression for the normal is
$\hat{\mathbf{n}}^{(1)}=\left[(2\pi A/\lambda)\sin(2\pi
  x/\lambda),0,0\right]$.
Solving for the charge density we obtain
$\sigma^{(1)}=-(\Delta\epsilon/\bar{\epsilon})(2\pi
A/\lambda)(Ze/4\pi)x\sin(2\pi x/\lambda)/(x^2+y^2)^{3/2}$.
The electric potential follows from the integration of the product of this
charge density and the Coulomb potential,
$(4\pi\epsilon_0)^{-1}(x^2+y^2)^{-1}$.
At the ion position, the leading correction term is
$\phi_\mathrm{P}^{(1)}=-(\Delta \epsilon/\bar{\epsilon}) A(Ze/4\pi\epsilon_0)(2\pi/\lambda)^2C|\ln(a/\lambda)|$,
with $C$ a positive constant.  For positive ions this excess potential is
negative.  At other positions, the leading correction to the potential is
approximately
$\phi_\mathrm{P}^{(1)}=-(\Delta \epsilon/\bar{\epsilon}) A(Ze/4\pi\epsilon_0)(2\pi/\lambda)^2C|\ln(a/\lambda)|\cos(2\pi x/\lambda)$.
We now use the fact that the mean curvature of the surface is
$H =- (1/2)(2\pi/\lambda)^2A\cos(2\pi x/\lambda)$. 
This curvature is negative in convex regions, such as those
around the peaks of the surface. Our result for the potential can be
expressed in terms of the curvature,
$\phi_\mathrm{P}^{(1)}=(\Delta \epsilon/\bar{\epsilon}) (Ze/4\pi\epsilon_0)(2CH)|\ln(a/\lambda)|$. 
We observe that this expression can be used as an approximation for the
potential in cases with a different modulation of the surface.
The resulting first-order contribution to the interaction energy 
between ion and surface is
$U^{(1)} = (\Delta \epsilon/\bar{\epsilon})
(Z^2e^2/4\pi\epsilon_0)(CH)|\ln(a/\lambda)|$.

For a particle near the surface the dominant contribution to its energy is its
interaction with the polarization charges. The ion distribution near the
surface therefore follows from the Boltzmann population factor
$\exp[-(U^{(0)}+U^{(1)})/k_\mathrm{B}T]$. Expanding the exponential factor and
multiplying by the bulk densities, we obtain the excess charge density near
the surface. Within an atomic diameter from the surface, the net accumulated
charge per unit area takes the approximate form
\begin{eqnarray}
  \delta q &=& - l_{\mathrm{B}} \left[C_1-C_2Aa 
    \frac{|\ln(a/\lambda)|}{\lambda^{2}}
    \cos\left(\frac{2\pi x}{\lambda}\right)\right] \nonumber\\
  && \times \sum_i c_i eZ_i^3 \;,
  \label{eq:LPB-density}
\end{eqnarray}
where $c_i$ is the bulk number density of species~$i$, and the integration
constants $C_1$ and $C_2$ are positive according to the functional form of the
estimated potentials. The values of the constants can be estimated in terms of
the parameters of the system but we note that, within the calculation outlined
above, they depend on the specific cutoff~$a$ chosen.
Equation~\eqref{eq:LPB-density} retains the dependence on valencies and
characteristic lengths. In particular, we emphasize that for asymmetric
electrolytes the result is nonzero. The net charge is a result of the
asymmetric depletion of ions near the interface. Additionally, the sign of the
first-order term indicates that the depletion is stronger at concave regions.
This result ignores ion correlations and is based on the properties of the
direct interaction of individual ions with the dielectric interface. Yet, as
shown below, it reproduces the key features of the charge distribution
observed in simulations, indicating that it likely represents the dominant
contribution.

\section{Results and Discussion}
\label{sec:results}

\begin{figure}
  \centering
  \includegraphics[width=\figurewidth]{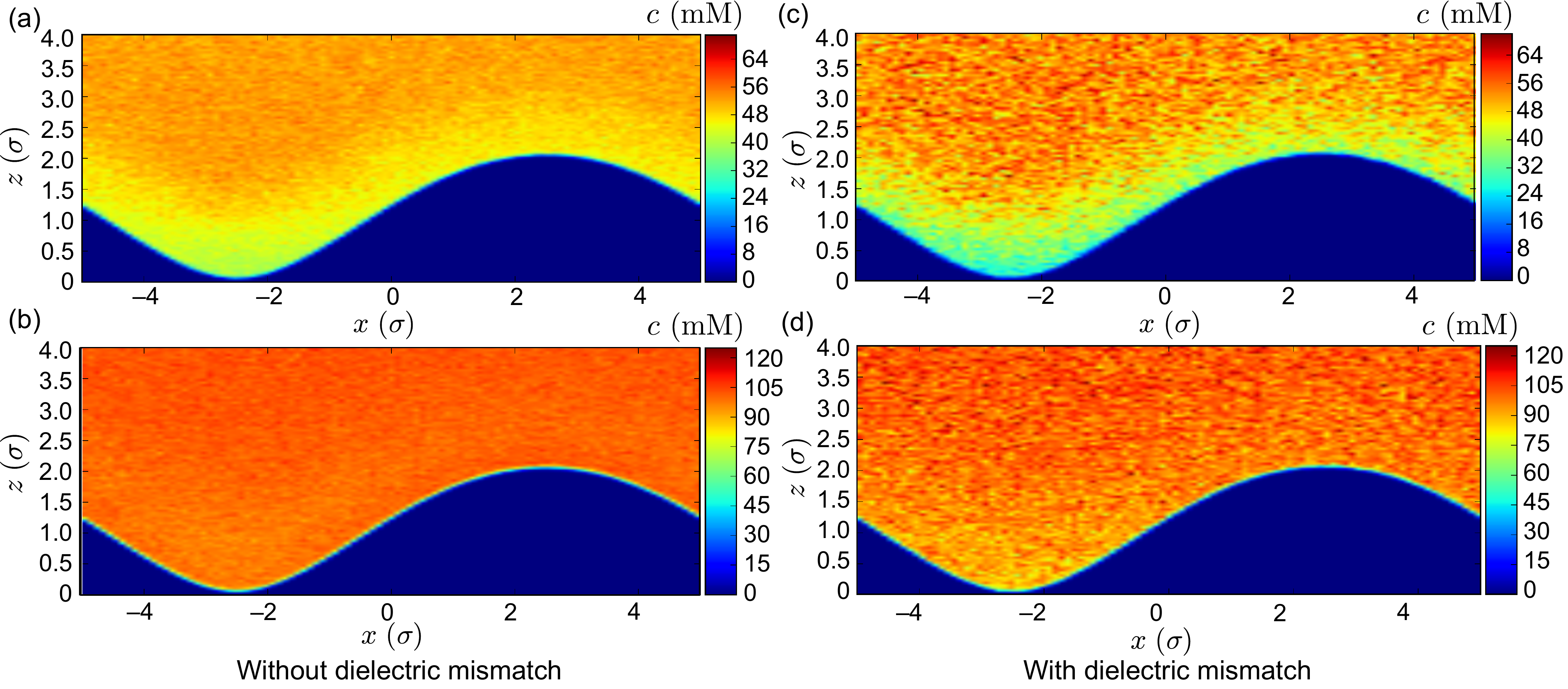}
  \caption{Density distributions of a 50~mM 2:1 electrolyte above a structured
    interface. Left: (a) Divalent and (b) monovalent ion number density
    distributions for a surface \emph{without} dielectric mismatch. Right:
    (c)~Divalent and (d)~monovalent ion density distributions for a surface
    with permittivity mismatch 80/2. The polarization charges significantly
    enhance the surface depletion, in particular for the divalent ions. All
    ion concentrations~$c$ are expressed in mM.}
\label{fig:density_dist}
\end{figure}

Figure~\ref{fig:density_dist} shows the ion number density for the 50~mM 2:1
electrolyte near the modulated surface.  In the absence of dielectric contrast
(left-hand panels in Fig.~\ref{fig:density_dist}) the bulk monovalent ion
density is almost exactly twice that of the divalent ions; only close to the
surface a small depletion occurs, which is more pronounced for the divalent
ions.  This effect appears as ions near the interface lack a symmetric shell
of screening counterions.  The asymmetric counterion shell pulls the ions
towards the bulk\cite{gan15}.  These results serve as a baseline to assess the
effects of the case with dielectric contrast.

\begin{figure}[b]
  \centering
  \includegraphics[width=\figurewidth]{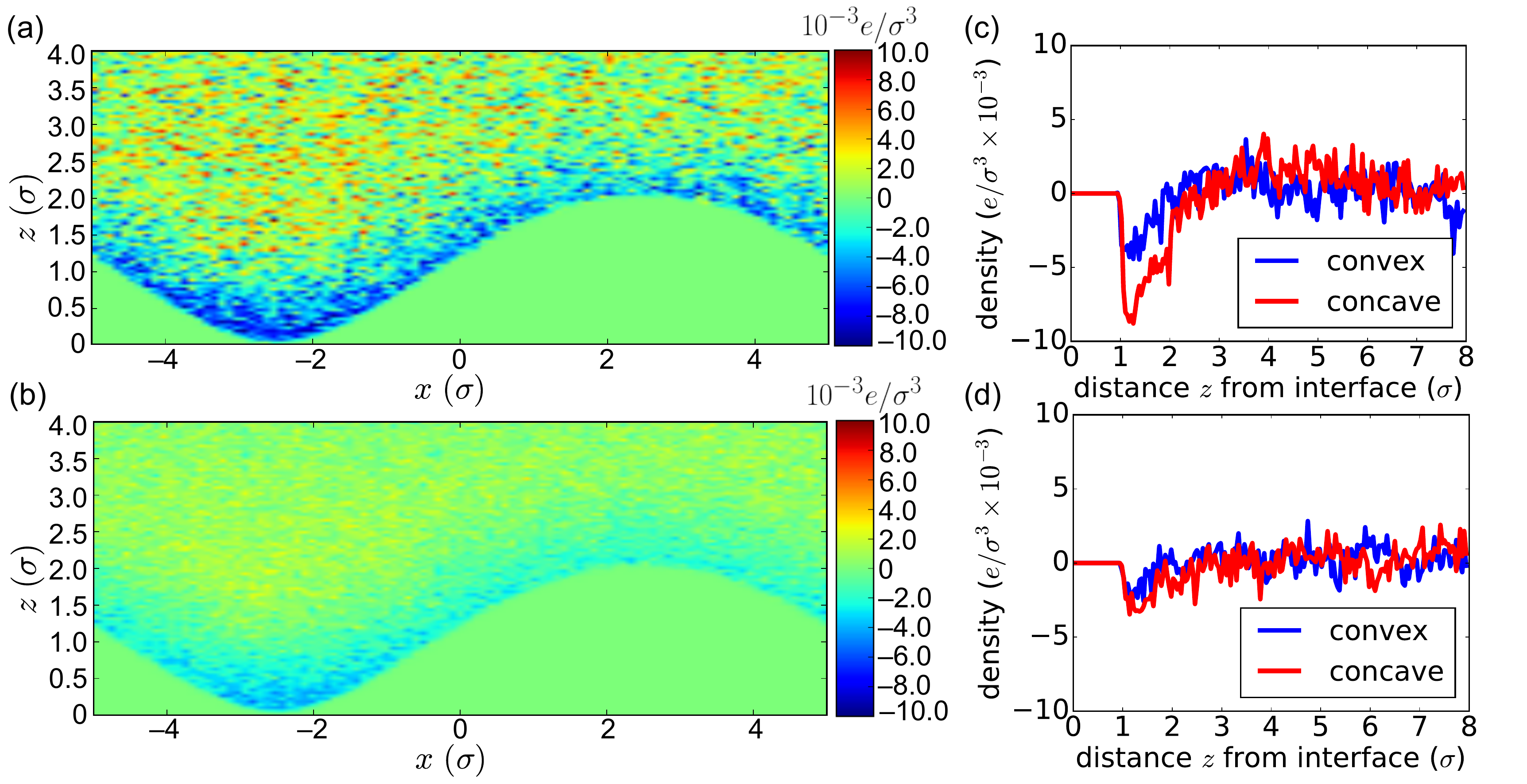}
  \caption{Net ionic charge distribution formed by 50~mM 2:1 electrolyte above
    a neutral structured interface with (a)~permittivity mismatch $80/2$ and
    (b)~no dielectric mismatch. The net ionic charge density is significantly
    enhanced by dielectric effects. In addition, the lateral position above
    the surface also affects the net charge density, as confirmed in panels
    (c) and~(d), with and without permittivity mismatch, respectively. The
    magnitude of the net charge density is largest above the concave regions
    (troughs) of the surface.  Panels (c) and~(d) were obtained using
    simulations based upon the variational approach of
    Ref.~\onlinecite{shen17}, and have a bin width of $0.5\sigma$ along the
    $x$ direction and $0.04\sigma$ in the $z$ direction.}
  \label{fig:charge_density}
\end{figure}

In the presence of dielectric contrast (right-hand panels in
Fig.~\ref{fig:density_dist}) we observe a stronger depletion of both charged
species, owing to the repulsive polarization charges.  The depletion now
extends further into the bulk, as can be expected from the long-range nature
of the electrostatic interactions.  More importantly, since the interaction
between the ion and its polarization charge scales as $Z^2$, the divalent ions
are significant more depleted near the surface than the monovalent ions. This
asymmetry breaks the concentration balance $2 c_{+2} = c_{-1}$ that is
fulfilled in the bulk, so that charge neutrality is violated in the vicinity
of the surface, with a net negative charge cloud above the surface
[Fig.~\ref{fig:charge_density}(a)]. Strictly, this effect also occurs in the
absence of dielectric mismatch [Fig.~\ref{fig:charge_density}(b)] owing to the
above-mentioned difference in asymmetry of the counterion shell, but the net
charge density is substantially stronger in the presence of dielectric
contrast.  Also, we observe that the depletion effect is stronger near concave
domains of the surface than near convex ones
[Fig.~\ref{fig:charge_density}(c)]. This result matches the modulation of the
excess charge found in our analytic calculation Eq.~\eqref{eq:LPB-density}.

\begin{figure}
  \centering
  \includegraphics[width=\figurewidth]{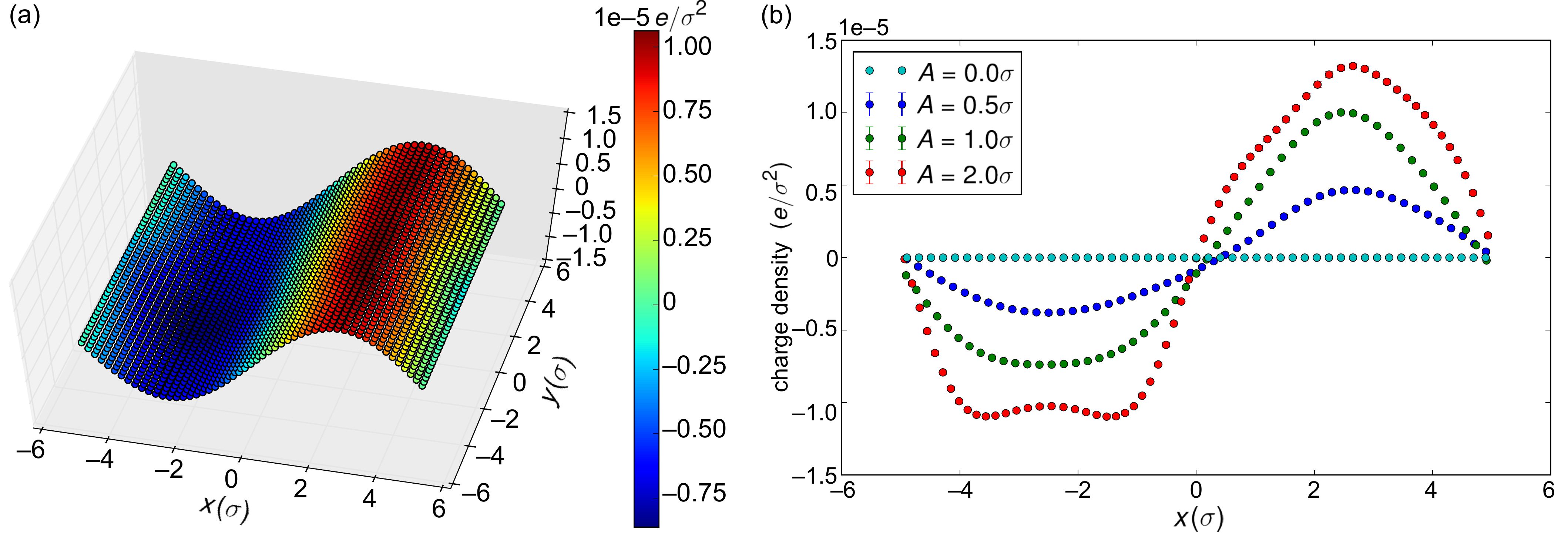}
  \caption{(a)~Time-averaged surface polarization charge density induced by
    the spatially modulated ion distribution.  (b)~Surface polarization charge
    profile for various amplitudes of the profile. The asymmetry near the
    maximum lies within the statistical accuracy.}
\label{fig:amp_variation}
\end{figure}

Along with the net ionic charge density in the electrolyte, the simulations
also provide the average \emph{induced} surface charge density.  Although
globally the net induced charge of the interface must vanish, it presents
persistent nonzero averages as a function of position. Consistent with the
modulation of the ionic charge density, the average induced charge density is
positive at convex regions and negative at the concave regions, as illustrated
by the time average in Fig.~\ref{fig:amp_variation}(a).

To further examine the dependence of the induced charge and ion charge density
on surface structure we systematically vary the parameters of the modulated
surface. We perform simulations for different modulation amplitudes~$A$
ranging from $0$ to $2\sigma$. Figure~\ref{fig:amp_variation}(b) shows the
induced charge density averaged over the $y$ direction, along which the
properties of the system are invariant.  For large amplitudes, we observe that
the induced charge density amplitude is larger and varies more rapidly at the
peak than in the trough.
At low amplitude, the induced charge density itself mimics the sinusoidal
variation of the surface, but this similarity breaks down at high amplitude
($A = 2.0\sigma$). This break-down reflects steric effects, where ions cannot
reach the bottom of the trough once the gap near the minimum becomes too
narrow.  It is noteworthy that asymmetries in ion size could further
complicate the observed density distributions. In particular, it is possible
that nonuniform ion distributions could even be reproduced for charge-symmetric
salts with size asymmetry between the anions and cations~\cite{messina02}.

\begin{figure}
  \centering
  \includegraphics[width=\narrowfigurewidth]{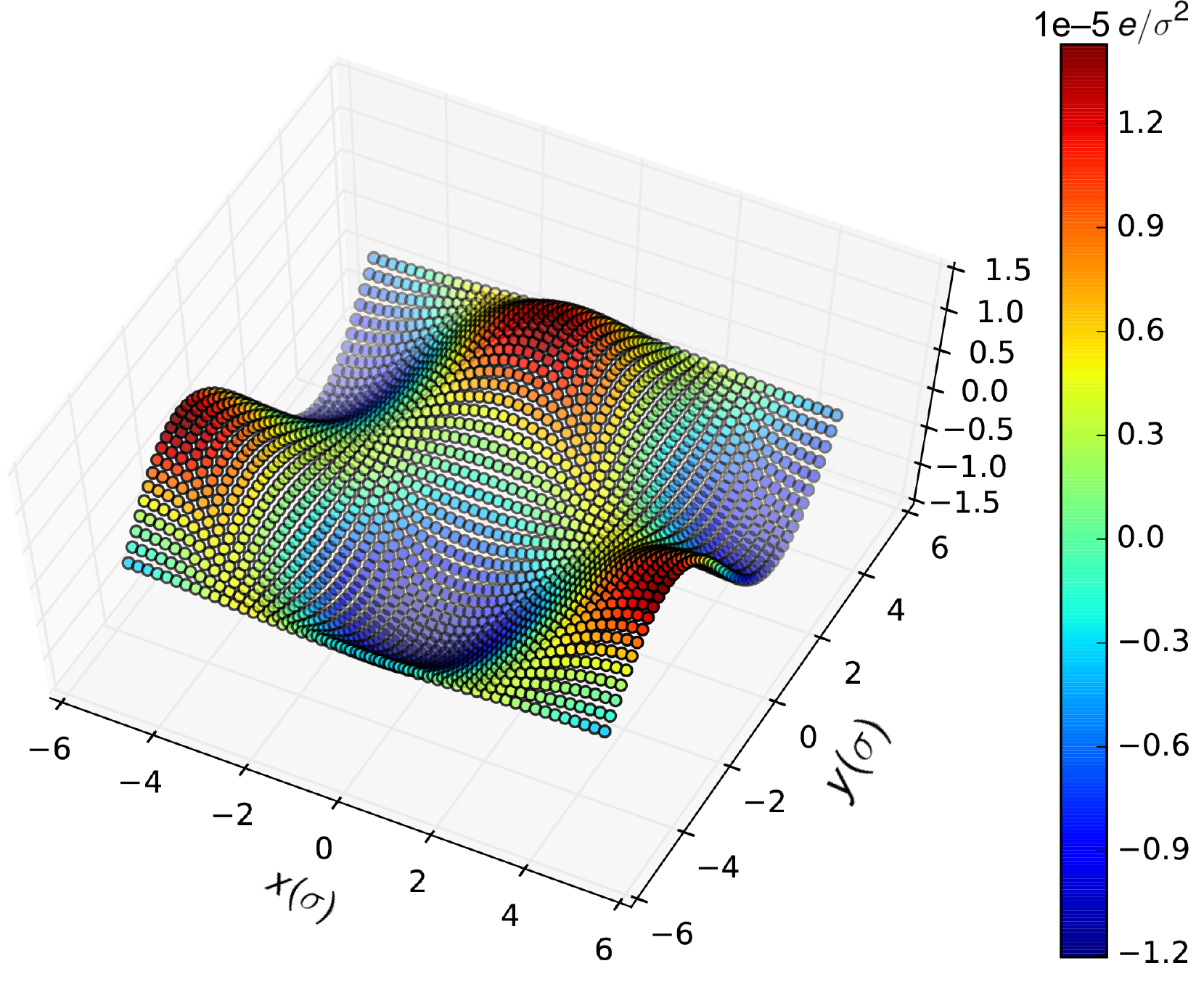}
  \caption{Induced surface charge density at a structured dielectric interface
    that is modulated along both the $x$ and $y$ directions.  As in the
    sinusoidal case [Fig.~\ref{fig:amp_variation}(a)], the dielectric mismatch
    is 80/2 with a 50~mM 2:1 electrolyte placed above the surface.}
\label{fig:egg_carton}
\end{figure}

Lastly, the phenomenon we have found in our simulations as well as the PB
analysis, of curvature-dependent charge depletion, is generic and not limited
to surfaces with modulation in a single dimension.  Indeed, it can be
generalized to other structures. For example, Fig.~\ref{fig:egg_carton}
illustrates the net surface polarization charge pattern of the same 50~mM 2:1
electrolyte above a structured dielectric interface with permittivity mismatch
80/2, but a surface modulation in both $x$ and $y$ directions,
$z(x, y) = A\cos(kx)\sin(ky)$.  Similar to our previous results, the valleys
of the surface acquire a negative surface polarization charge, whereas the
peaks carry a positive induced charge. We note that periodicity of the
modulation is not a requirement for the phenomenon to occur.

The simulations presented, along with the arguments based on single-ion
interactions with the surface, demonstrate that the effect observed is
universal.  The local curvature of the surface always induces effective
surface polarization and net ion charge accumulation in the presence of
asymmetric electrolytes.  The effect should be observable not only on surfaces
that bound an electrolyte, but also at the surface of electrolyte-immersed
colloids. Our findings can be applied to the design of surfaces with useful
physical--chemical properties.

\begin{acknowledgments}
  This research was supported through award 70NANB14H012 from the U.S.
  Department of Commerce, National Institute of Standards and Technology, as
  part of the Center for Hierarchical Materials Design (CHiMaD), the National
  Science Foundation through Grant No.\ DMR-1610796, and the Center for
  Computation and Theory of Soft Materials (CCTSM) at Northwestern
  University. We thank the Quest high-performance computing facility at
  Northwestern University for computational resources.
\end{acknowledgments}

\end{document}